\documentclass{article}

\usepackage{microtype}
\usepackage{graphicx}
\usepackage{subfigure}
\usepackage{booktabs} 

\usepackage{amsmath}
\usepackage{amssymb}
\usepackage{caption}

\usepackage{graphicx}
\usepackage{multirow}
\usepackage{tabularx}

\newcolumntype{M}[1]{>{\centering\arraybackslash}m{#1}}
\newcolumntype{R}[1]{>{\raggedleft\arraybackslash}m{#1}} 
\newcolumntype{L}[1]{>{\raggedright\arraybackslash}m{#1}} 

\newcommand{\ra}[1]{\renewcommand{\arraystretch}{#1}}

\usepackage{enumitem}

\usepackage{hyperref}

\usepackage[accepted]{mlsys2021}

\mlsystitlerunning{MicroRec: Efficient Recommendation Inference by Hardware and Data Structure Solutions}

\begin{document}

\twocolumn[

\begin{center}
\vspace{-5em}
\large{This paper has been accepted by the 4th Conference on Machine Learning and Systems (MLSys'21)}
\vspace{2em}
\end{center}

\mlsystitle{MicroRec: Efficient Recommendation Inference \\by Hardware and Data Structure Solutions}



\mlsyssetsymbol{equal}{*}

\begin{mlsysauthorlist}
\mlsysauthor{Wenqi Jiang}{eth,cuni}
\mlsysauthor{Zhenhao He}{eth}
\mlsysauthor{Shuai Zhang}{eth}
\mlsysauthor{Thomas B. Preußer}{eth}
\mlsysauthor{Kai Zeng}{ali}
\mlsysauthor{Liang Feng}{ali}
\mlsysauthor{Jiansong Zhang}{ali}
\mlsysauthor{Tongxuan Liu}{ali}
\mlsysauthor{Yong Li}{ali}
\mlsysauthor{Jingren Zhou}{ali}
\mlsysauthor{Ce Zhang}{eth}
\mlsysauthor{Gustavo Alonso}{eth}
\end{mlsysauthorlist}

\mlsysaffiliation{eth}{ETH Zurich}
\mlsysaffiliation{ali}{Alibaba Group}
\mlsysaffiliation{cuni}{Columbia University}

\mlsyscorrespondingauthor{Wenqi Jiang}{wenqi.jiang@inf.ethz.ch}

\mlsyskeywords{MLSys, Recommender Systems, Hardware Accelerator, FPGA}

\vskip 0.3in

\begin{abstract}
{
Deep neural networks are widely used in personalized recommendation systems. Unlike regular DNN inference workloads, recommendation inference is memory-bound due to the many random memory accesses needed to lookup the embedding tables. The inference is also heavily constrained in terms of latency because producing a recommendation for a user must be done in about tens of milliseconds. In this paper, we propose MicroRec, a high-performance inference engine for recommendation systems. MicroRec accelerates recommendation inference by (1) redesigning the data structures involved in the embeddings to reduce the number of lookups needed and (2) taking advantage of the availability of High-Bandwidth Memory (HBM) in FPGA accelerators to tackle the latency by enabling parallel lookups. 
We have implemented the resulting design on an FPGA board including the embedding lookup step as well as the complete inference process. Compared to the optimized CPU baseline (16 vCPU, AVX2-enabled), MicroRec achieves 13.8$\sim$14.7$\times$  speedup on embedding lookup alone and 2.5$\sim$5.4$\times$ speedup for the entire recommendation inference in terms of throughput. As for latency, CPU-based engines needs milliseconds for inferring a recommendation while
MicroRec only takes microseconds, a significant advantage in real-time recommendation systems.
}
\end{abstract}
]



\printAffiliationsAndNotice{}  

\section{Introduction}
\label{sec:intro} 
\vspace{-0.5em}
 
Personalized recommendations are widely used to improve user experience and increase sales. Nowadays, deep learning has become an essential building block in such systems. For example, Google deploys wide-and-deep models for video and application recommendations~\cite{wide_and_deep_app_store, wide_and_deep_MT_youtube}; Facebook uses different kinds of deep models for a range of social media scenarios~\cite{facebook_benchmark}; and Alibaba combines attention mechanism with DNNs and RNNs for online retail recommendations~\cite{din_alibaba_attention_fc, dien_alibaba_attention_rnn}. Due to the popularity of DNN-based recommendation models, they can comprise as much as 79\% of the machine learning inference workloads running in data centers~\cite{facebook_benchmark}.

\begin{figure*}[t]
  \centering
  \includegraphics[width=0.9\linewidth]{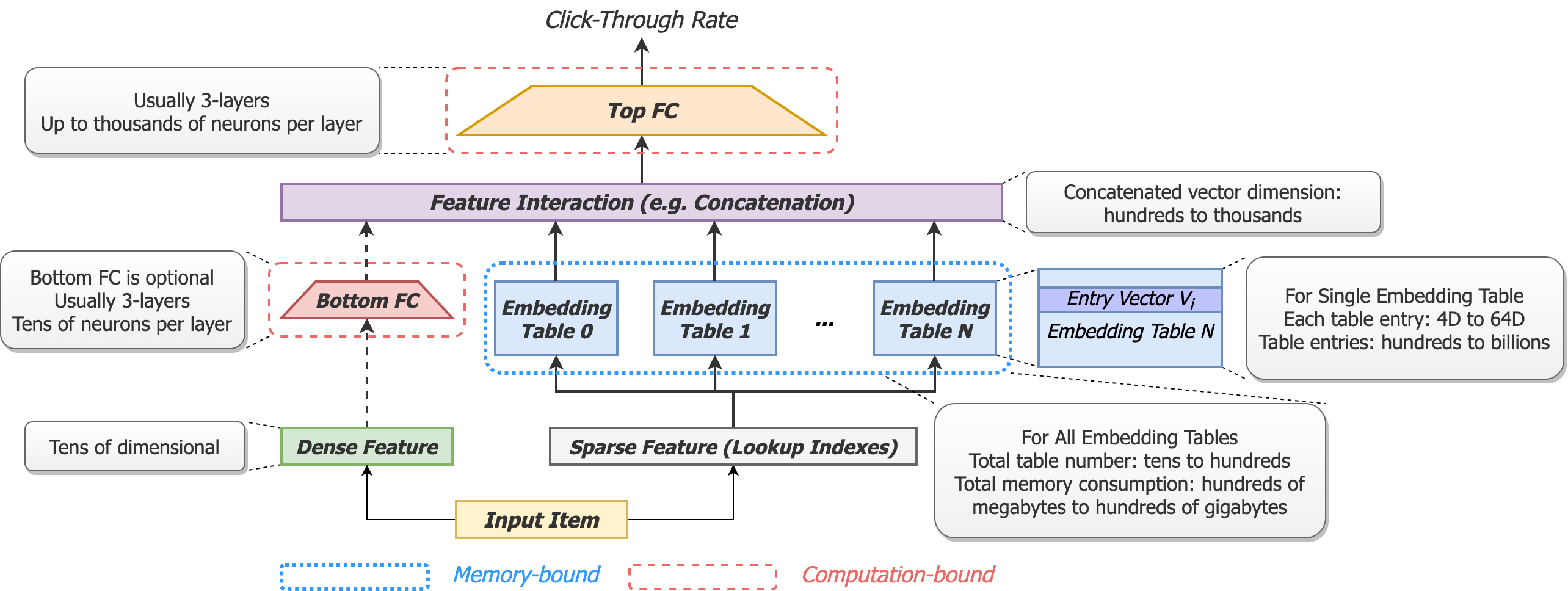}
  \vspace{-0.7em}
  \caption{A typical deep recommendation model and it's workload specification.}
  \label{fig:embedding}
\end{figure*}

\begin{figure*}[t]
  \centering
  \includegraphics[width=\linewidth]{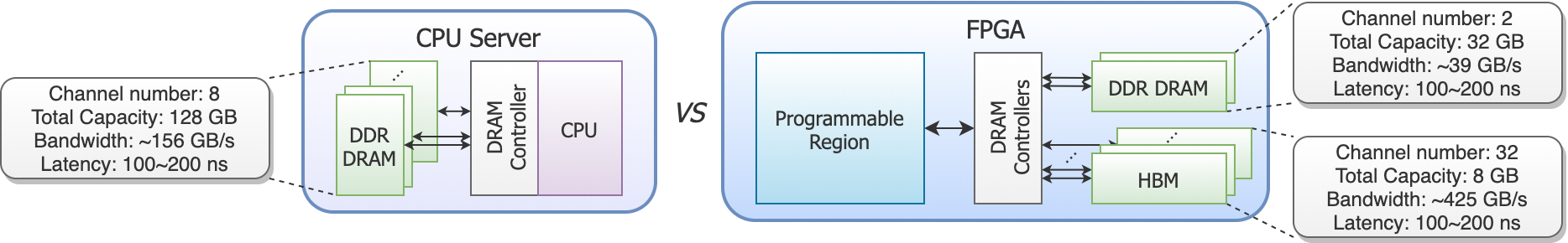}
  \vspace{-2em}
  \caption{Two hardware choices for recommendation inference. Left: a typical CPU server on which models are stored in DDR DRAM (memory channel number varies from server to server) and computation is done in CPU. Right: our FPGA accelerator where embedding tables are distributed over many memory channels and fast inference is supported by reprogrammable circuit.}
  \vspace*{-3mm}
  \label{fig:fpga_cpu_system}
\end{figure*}

\vspace{-1em}
\paragraph*{Deep Recommendation Models} We first briefly introduce deep recommendation models to provide the necessary context to discuss the challenges, our methods, and contributions.
Figure~\ref{fig:embedding} illustrates a classical deep recommendation model for \textit{Click-Through Rate} (CTR) prediction~\cite{facebook_benchmark, wide_and_deep_app_store} and summarize its workload characteristics. An input feature vector consists of dense features (e.g., age and gender) and sparse features (e.g., location and advertisement category). Over the dense feature
vector, some systems apply a neural feature extractor that consists of multiple fully connected (FC) layers~\cite{facebook_benchmark, tensordimm_kaist}, while some design~\cite{wide_and_deep_app_store} does not contain the bottom FC layers. Over the sparse feature vector, the system translates each feature into a dense feature embedding by looking up it in an embedding table. These features are then combined (e.g., concatenated) and fed to a neural classification model consisting of multiple fully connected layers.

\vspace{-1em}
\paragraph*{Challenges in a CPU-based System} 
When deploying recommendation systems on typical CPU servers (left half of Figure~\ref{fig:fpga_cpu_system}), embedding tables are stored in DDR DRAM, and the cores are responsible for the computation. 
There are two system bottlenecks in such deployments.

\textit{First}, embedding table lookups are costly because they induce massive random DRAM accesses on CPU servers.
Production recommendation models usually consist of at least tens of embedding tables, thus each inference requires the corresponding lookup operations. Due to the tiny size of each embedding vector, the resulting DRAM accesses are nearly random rather than sequential. Since CPU servers have only a few memory channels, these random DRAM accesses are expensive. 

\textit{Second}, both embedding lookups and computation can be expensive if one resorts to ML frameworks such as TensorFlow and PyTorch. According to our observations, on \textit{TensorFlow Serving} which is optimized for inference, the embedding layer involves 37 types of operators (e.g., concatenation and slice) and these operators are invoked multiple times during inference,
resulting in significant time consumption especially in small batches. 
Similarly, the throughput of neural network computation can also be restricted when using small batches.
Unfortunately, small batch sizes are usually required in CPU-based recommendation engines to meet the latency requirements of tens of milliseconds, thus the framework overhead is non-negligible.

Not surprisingly, there has been a range of work trying to accelerate deep recommendation models. \citet{tensordimm_kaist} and \citet{facebook_benchmark} observed the main system bottleneck of substantial random memory accesses. \citet{tensordimm_kaist} and \citet{ke2020recnmp} thus proposed to redesign DRAM in micro-architectural level; however, it would take years to put such new DRAM chips in production even if they are adopted.
\citet{gupta2020deeprecsys} suggested GPUs could be useful in recommendation for large batches, but the memory bottleneck still remains and GPUs suffer from high latency. Similarly, \citet{hwang2020centaur} implemented an FPGA accelerator for recommendation but without removing the memory bottleneck.
In this paper, we ask: \textit{Can we
accelerate deep recommendation models, at industrial scale, with \textit{practical} yet \textit{efficient} hardware acceleration?}

\paragraph*{Our Approach}
Based on careful analysis of two production-scale models from Alibaba, we design and implement MicroRec, a low-latency and high-throughput recommendation inference engine. Our speed-ups are rooted in two sources. 
First, we employ more suitable hardware architecture for recommendation with (a) hybrid memory system containing High Bandwidth Memory (HBM), an emerging DRAM technology, for highly concurrent embedding lookups; and (b) deeply pipelined dataflow on FPGA for low-latency neural network inference. 
Second, we revisit the data structures used for embedding tables to reduce the number of memory accesses. By applying Cartesian products to combine some of the tables, the number of  DRAM accesses required to finish the lookups are significantly reduced .

Our contributions in this paper include:
\begin{enumerate}[wide = 0pt]
	\item We show how to use high-bandwidth memory to scale up the concurrency of embedding lookups. This introduces 8.2$\sim$11.1$\times$ speedup over the CPU baseline.
	\item To the best of our knowledge, this is the first paper that proposes to reduce the number of random memory accesses in deep recommendation systems by data structure design. We show that applying Cartesian Products between embedding tables further improves the lookup performance by 1.39$\sim$1.69$\times$ with marginal storage overhead (1.9$\sim$3.2\%). 
	\item To optimize performance with low storage overhead, we propose a heuristic algorithm to combine and allocate tables to the hybrid memory system on the FPGA.
    \item We implement MicroRec on FPGA and test it on two production models from Alibaba  (47 tables, 1.3 GB; 98 tables, 15.1 GB). The end-to-end latency for a single inference only consumes 16.3$\sim$31.0 microseconds, 3 to 4 orders of magnitude lower than common latency requirements for recommender systems. In terms of throughput, MicroRec achieves 13.8$\sim$14.7$\times$  speedup on the embedding layer, and 2.5$\sim$5.4$\times$ speedup on the complete inference process compared to the baseline (16 vCPU; 128 GB DRAM with 8 channels; AVX2-enabled). 
\end{enumerate}

\section{Deep recommendation Systems}
\label{sec:deep_rec}

Personalized recommendation systems are widely deployed by YouTube~\cite{deep_youtube, wide_and_deep_MT_youtube}, Netflix~\cite{gomez2015netflix}, Facebook~\cite{park2018deep}, Alibaba~\cite{din_alibaba_attention_fc, dien_alibaba_attention_rnn}, and a number of other companies~\cite{underwood2019use, xie2018personalized, chui2018notes}. 
In this section, we review their basic properties and analyze their performance to identify the main bottlenecks.

\subsection{Deep Model for Ranking}
\label{sec:deep_ranking}

Figure \ref{fig:embedding} abstracts the deep model for recommendation ranking that we target: it is responsible for predicting \textit{click-through-rates} (CTR), i.e., how likely it is that the user will click on the product. 
The model takes a set of sparse and dense features as input. For example, account IDs and region information are encoded as one-hot vector (sparse feature), while age serves as part of the dense feature since the number is consecutive. 
The prediction process is as follows. First, dense and sparse input features are processed separately. Depending on the model design, dense features can be processed by a few fully-connected layers~\cite{naumov2019dlrm} or served \textit{as-is} without any pre-processing~\cite{wide_and_deep_app_store, din_alibaba_attention_fc}. The sparse features, on the other hand, are converted to a set of indexes to lookup vectors from embedding tables. For each inference task, one or several vectors are retrieved from each table~\cite{facebook_benchmark}. The embedding vectors so retrieved are then concatenated with raw or processed dense features. 
Finally, the concatenated vectors are fed to the top fully-connected layers for CTR prediction. Product candidates with the highest CTRs are recommended to users.

The specific model design varies from scenario to scenario. Some adjustable parameters include: number of fully-connected layers, number of hidden neurons in each layer, numbers and sizes of embedding tables, feature interaction operations (e.g., concatenation, weighted sum, and element-wise multiplication), whether to include bottom fully-connected (FC) layers.\footnote{The models we target do not contain bottom FCs, and each table is looked up only once.}

\subsection{Embedding Table Lookups}
\label{sec:embedding}

Embedding table lookup is the key difference between deep recommendation models and regular DNN workloads, and it shows the following traits. First, the embedding tables contribute to the majority of storage consumption in deep recommendation models. Large embedding tables at industry scale can contain up to hundreds of millions of entries, consuming tens or even hundreds of gigabytes of storage. 
Second, the size of the tables varies wildly between a few hundred (e.g., countries or ``province ID'') to hundreds of millions of entries (e.g., ``user account ID'').

Embedding table lookup is problematic from a performance perspective. Due to the traits mentioned above, most tables are held in main memory, inducing many random memory accesses during inference. \citet{ke2020recnmp} proves this point by showing that high cache miss rates are common in deep recommendation inference.

\begin{figure}[t]
	\centering
  \includegraphics[width=1.0\linewidth]{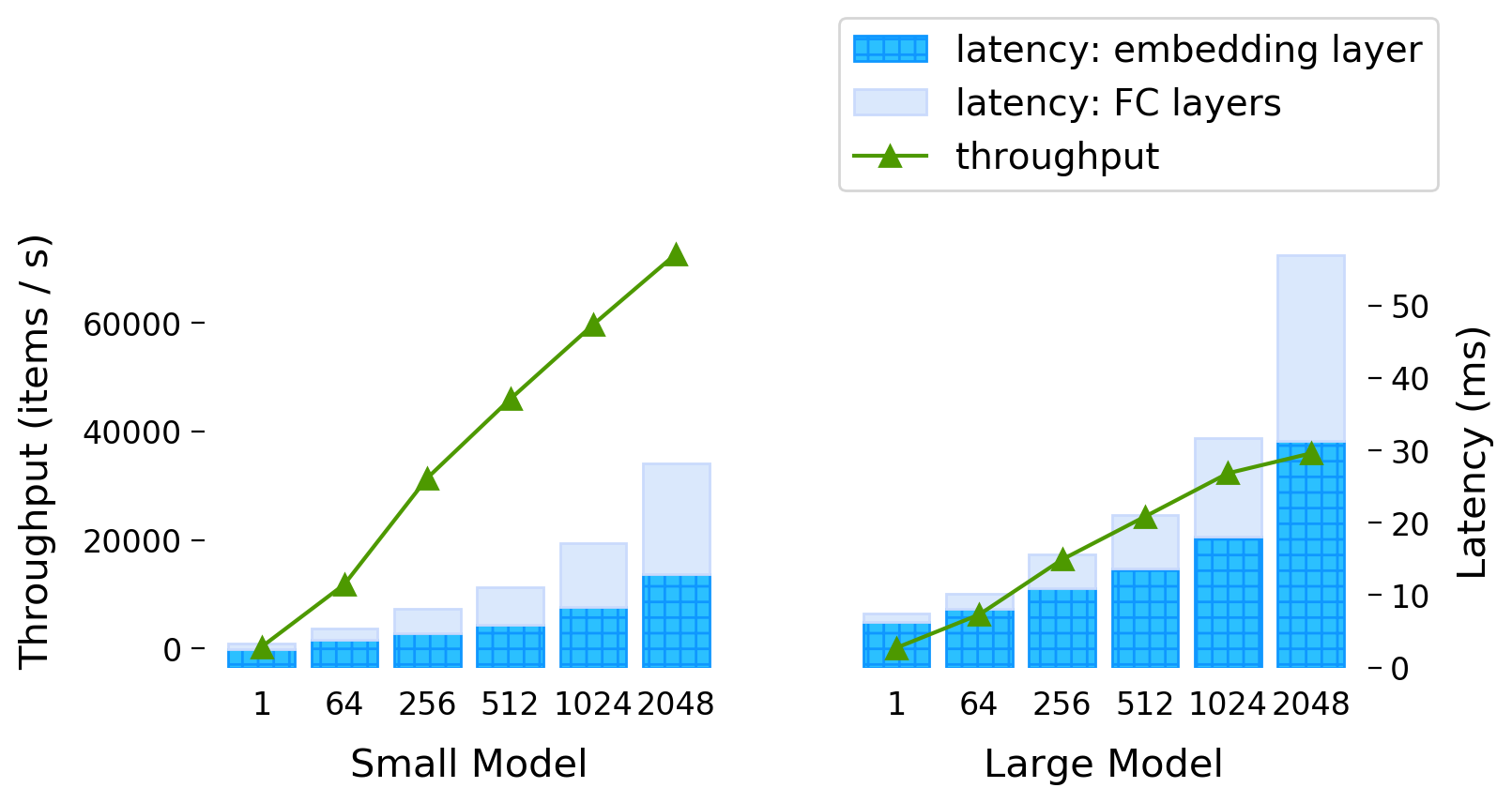}
  \vspace*{-9mm} 
  \caption{The embedding layer is expensive during inference.}
  \label{fig:cpu_perf_breakdown}
\end{figure}

\subsection{Performance Analysis}
\label{sec:cpu_perf}

We chose CPUs as the hardware platform for baseline experiments. 
Although GPUs are popular for neural network training, they have not shown clear advantages over CPUs for deep recommendation inference. As reported by \citet{gupta2020deeprecsys}, GPUs can only outperform CPUs when (a) the model is computation-intensive (less embedding lookups), and (b) very large batch sizes are used. 

Figure \ref{fig:cpu_perf_breakdown} shows the cost of the embedding layer during inference on two models from Alibaba (models specified in Table \ref{tab:model_size}) .
As a side effect of the massive number of memory accesses, the many related operators also lead to significant overhead. According to our observation on \textit{TensorFlow Serving}, an optimized ML framework for inference, 37 types of operators are involved in the embedding layer (e.g., slice and concatenation), and these operators are invoked many times during inference. The close latency to infer small batches (size of 1 and 64) illustrates the expense of operator-calls. 
Larger batch sizes can lead to better throughput, yet SLA (latency requirement) of tens of milliseconds must be met, thus extremely large batches are not allowed for recommendations. 

\section{MicroRec}
\label{sec:microrec}

We present MicroRec, an FPGA-enabled high-performance recommendation inference engine which involves both hardware and data structure solutions to reduce the memory bottleneck caused by embedding lookups. On the hardware side, our FPGA accelerator features highly concurrent embedding lookups on a hybrid memory system (HBM, DDR DRAM, and on-chip memory). On the data structure side, we apply Cartesian products to combine tables so as to reduce random memory accesses. Putting them together, we show how to find an efficient strategy to combine tables and allocate them across hybrid memory resources.

\begin{figure}[t!]
	\centering
  \includegraphics[width=1.0\linewidth]{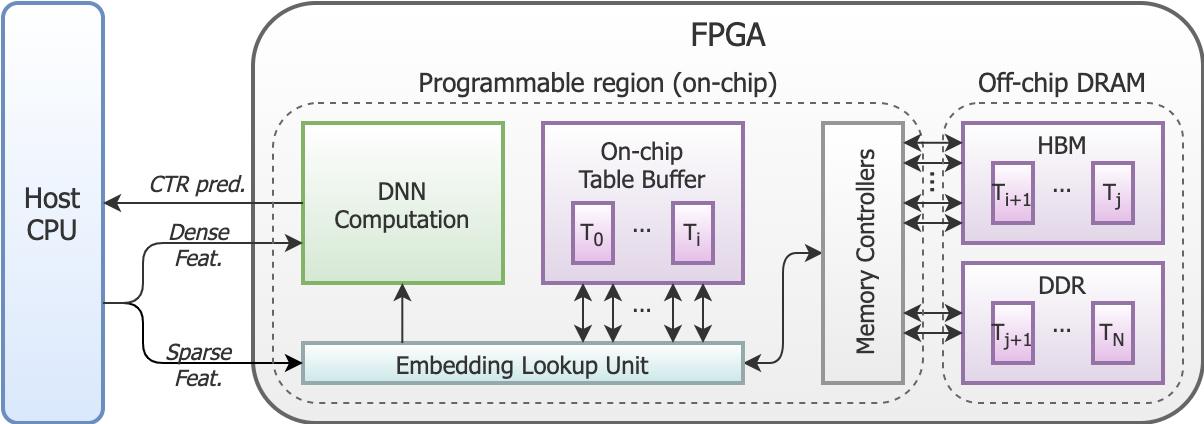}
  \vspace{-2em} 
  \caption{System overview of MicroRec.
  }
  \label{fig:system_overview}
  \vspace{-2em}
\end{figure}

\subsection{System Overview}
\label{sec:microrec_overview}

Figure \ref{fig:system_overview} overviews the hardware design of MicroRec. Embedding tables are distributed over both on-chip memory (BRAM and URAM) and off-chip memory (HBM and DDR). Neural network inference is taken cared by the DNN computation units which contain both on-chip buffers storing weights of the model and computation resources for fast inference. 
To conduct inference, the host server first streams dense and sparse features to the FPGA\footnote{\label{foot_CPU_input}The Vitis hardware development platform does not yet support streaming from the host server to a Xilinx U280 FPGA, thus we have prototyped the design by caching the input features on FPGA.}. Then, the embedding lookup unit translates the sparse features to dense vectors by looking up embedding tables from both on-chip and off-chip memory. Finally, the computation unit takes the concatenated dense vector as input and finishes inference before returning the predicted CTR to the host. 

\subsection{Boost Emebdding Lookup Concurrency by Increased Memory Channels}
\label{sec:microrec_memory}

The tens of embedding table lookup operations during inference can be parallelized when multiple memory channels are available. MicroRec resorts to high-bandwidth memory as the main force supporting highly concurrent embedding lookups. Besides that, we also take advantage of other memory resources on FPGA, i.e., DDR4 DRAM and on-chip memory, to further improve lookup performance. 

\subsubsection{High-Bandwidth Memory}
\label{sec:microrec_HBM}

We resort to HBM to parallelize embedding lookups. As an attractive solution for high-performance systems, HBM offers improved concurrency and bandwidth compared to conventional DRAMs~\cite{jun2017hbm, o2014highlights}. In this paper, we use a Xilinx Alveo U280 FPGA card~\cite{xilinx_u280} equipped with 8 GBs of HBM which provides a bandwidth of up to 425 GB/s~\cite{zeke2020benchmark_hbm}. More specifically, the HBM system on U280 consists of 32 memory banks, which can be accessed concurrently by independent pseudo-channels. Thus, embedding tables can be distributed to these banks so that each bank only contains one or a few tables, and up to 32 tables can be looked up concurrently.

\subsubsection{Hybrid Memory System on FPGA}
\label{sec:microrec_hybrid_memory}

The Xilinx Alveo U280 FPGA involves multiple types of memory resources, including on-chip memory (BRAM and URAM) and off-chip memory (DDR4 DRAM and HBM), which exhibit different traits. 
HBM and DDR show close access latency of a couple of hundreds of nanoseconds given the memory controller generated by Vitis~\cite{vitis}, but have different concurrency-capacity trade-off (HBM: 32 channels, 8GB; DRAM: 2 channels, 32 GB).
Besides HBM and DDR, FPGAs also equip a few megabytes of on-chip memory that plays a similar role as CPU cache (small yet fast memory to cache frequently-accessed data or intermediate results).
Without read initiation overhead as in DRAM, the latency to access on-chip memory only consists of control logic and sequential read. According to our experiments, finish retrieving an embedding vector from an on-chip memory bank only consumes up to around 1/3 time of DDR4 or HBM.

\subsection{Reduce Memory Accesses by Cartesian Products}
\label{sec:microrec_cartesian}

\begin{figure}
  \centering
    \includegraphics[width=1.0\columnwidth]{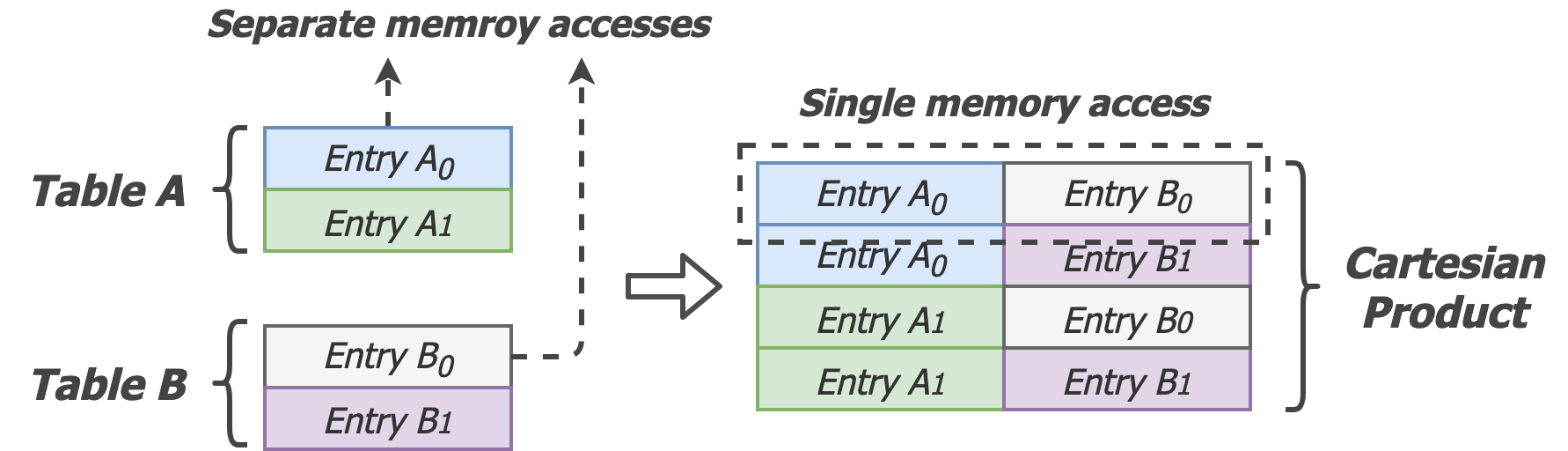}
    \vspace{-1em}
    \caption{Cartesian product of two embedding tables. Each entry of the product concatenates an entry from table A and another from B: one memory access retrieves two embedding vectors.}
    \vspace{-1em}
    \label{fig:cartesian}
\end{figure}

We reduce the number of memory accesses by combining tables so that each memory access can retrieve multiple embedding vectors. As shown in Figure \ref{fig:cartesian}, two embedding tables can be joined into a single larger one through a relation Cartesian Product. Since tables A and B in Figure \ref{fig:cartesian} have two entries, the product table ends up with four entries: each of them is a longer vector obtained by concatenating an entry from table A and another from table B.
Using such a representation, the number of memory accesses is reduced by half: instead of two separate lookup operations now only one access is needed to retrieve the two vectors. 

By applying a Cartesian product, the latency to lookup two tables is reduced by almost half. Embedding tables in deep recommendation models usually contain short entry vectors (with between 4 to 64 elements in most cases). Although the entry vectors of the product are longer, i.e., the sum of two individual entries, they are still not long enough to fully take advantage of the spatial locality within DRAM. To retrieve a vector up to a few hundreds of bytes, a DRAM spends most of the time initiating the row buffer, while the following short sequential scan is less significant in terms of time consumption. As a result, reducing the memory accesses by half can lead to a speedup of almost 2$\times$.

Though Cartesian products lead to higher storage consumption, this overhead is comparatively small. This may sound counter-intuitive, 
however, most deep recommendation models contain tables of different size scales, so applying Cartesian products on small tables is almost for free compared to some of the largest tables in the model. 
According to our observations of real-world deployments, while some tables only consist of 100 4-dimensional embedding vectors, large tables can contain up to hundreds of millions of entries with a vector length of 64 due to the reasons discussed in section \ref{sec:embedding}. In this case, a Cartesian product of two small tables requires only tens of kilobytes (assume 32-bit floating-point storage): almost negligible compared to a single large table of tens or hundreds of gigabytes.

Cartesian products can help balancing the workload on off-chip DRAM (DDR and HBM). 
For example, suppose there are 34 off-chip memory channels (32 for HBM and 2 for DDR), and 40 tables should be allocated on them. 
In this case, some banks have to store two embedding tables while others only hold one. 
When retrieving one vector from each table, the lookup performance is bound by the channels holding two tables, as the lookup latency on them is potentially 2$\times$ that of those containing only one table. 
Using Cartesian products, the total number of tables can be reduced from 40 to 34. This allows us to balance the workload on each memory channel resulting in potentially 2$\times$ speedup compared to an unbalanced workload situation. 

\subsection{Putting Everything Together: A Rule-based Algorithm for Table Combination and Allocation}
\label{sec:microrec_algorithm}

Our objective is to minimize embedding lookup latency given the memory constraints discussed in section \ref{sec:microrec_hybrid_memory}, i.e., available capacity and channels of each type of memory. To achieve this, an algorithm is required to explore solutions of combining tables through Cartesian products and deploying the result on memory banks.

\subsubsection{Brute-force Search}
\label{sec:microrec_brute_force}

A straightforward way to achieve this objective is to explore all possibilities in a brute-force manner and choose the best solution. 
First, one would list all possibilities of using tables as Cartesian product candidates. Then, for each one of these options, all possible combinations of Cartesian products would be calculated (including joining more than two tables). Based on the combinations of tables available, the single and combined tables are allocated to memory banks (solutions exceeding the memory capacity of a bank can be dropped) minimizing the latency. For ties in latency, the solution with the least storage overhead is chosen.

However, applying brute-force search is unrealistic because of the large exploration space. 
For example, selecting $n$ of out $N$ total tables as Cartesian candidates is a combinatorial problem with a time complexity of \(O(\frac{N!}{n!(N-n)!})\). Then, it costs \(O(n!)\) to explore any Cartesian products combinations of the candidates. Each outcome, including Cartesian products and original tables, are then allocated to memory banks at the cost of \(O(N)\). Using a parameter to control how many tables are selected for Cartesian products, the overall time complexity of the brute-force search is \(O(\sum_{n=1}^{N}N\frac{N!}{(N-n)!})\), making brute-force searching infeasible as the number of tables grows up.

\subsubsection{Heuristic-rule-based Search}
\label{sec:microrec_heuristic}
To optimize embedding lookup latency, we propose a heuristic search algorithm that can efficiently search for near-optima solutions with a low time complexity of $\mathcal{O}(N^2)$. Besides, this algorithm can be generalized to any FPGAs, no matter whether they are equipped with HBM, and no matter how many memory channels they have. Due to the memory traits introduced in section \ref{sec:microrec_hybrid_memory}, the algorithm simply regards HBM as additional memory channels: designers can adjust the memory channel number and bank capacities in the algorithm according to the available hardware.

Four heuristics are applied in the algorithm to reduce the search space where the optimal solution is unlikely to appear\footnote{The rules can be expanded, modified, or removed to adpat different models since these rules are table-size-dependent.}. Consequently, the algorithm can return near-optimas with low time complexity. The first three rules are designed to explore Cartesian combinations efficiently, while the fourth rule is for memory allocation.


Heuristic rule 1: large tables are not considered for Cartesian products. Tables are sorted by size and only the $n$ smallest tables should be selected for Cartesian products, otherwise products of large tables can lead to heavier storage overhead.

Heuristic rule 2: Cartesian products for table pairs of two. Although Cartesian products of the three smallest tables may only consume tens of megabytes storage (still small compared to a single large table of several or tens of gigabytes), the overall solution could be sub-optimal because this method consumes too many small tables at once while they are appropriate candidates to pair with larger tables.

Heuristic rule 3: within the product candidates, the smallest tables are paired with the largest tables for Cartesian products. This rule avoids terrible solutions where a Cartesian product is applied between two large tables. 

Heuristic rule 4: cache smallest tables on chip. 
After applying Cartesian products, we sort all tables by sizes and decide the number of small tables to store on chip.
Two constraints must be considered during this process. First, the size of selected tables should not exceed assigned on-chip storage. Second, if multiple tables are co-located in the same on-chip bank, the total lookup latency should not exceed off-chip (DDR or HBM) lookups, otherwise caching tables on-chip is meaningless.

Algorithm~\ref{algo:heuristic} sketches the heuristic-rule-based search for table combination and allocation. It starts by iterating over the number of tables selected as Cartesian product candidates. Within each iteration, the candidates are quickly combined by applying the first three heuristic rules ($\mathcal{O}(N)$). All tables are then allocated to memory banks efficiently by rule 4 ($\mathcal{O}(N)$). The algorithm ends by returning the searched solution that achieves the lowest embedding lookup latency. Considering the outer loop iterating over Cartesian candidate numbers, the total time complexity of the heuristic algorithm is as low as $\mathcal{O}(N^2)$.

\begin{algorithm}[t!]
\caption{Heuristic Search}
\label{algo:heuristic} 
\begin{algorithmic}
    \STATE {\bfseries Input}:
    $N$: total number of embedding tables; 
    $n$ : number of tables that are selected for Cartesian products; $c$: candidate tables for Cartesian products; $p$ : all tables after applying Cartesian products
    
    \STATE {\bfseries Output}: 
    $current\_best$: the best solution found by the algorithm, including the resulting table number and sizes as well as which banks they are allocated to.

    \vspace{-0.3em}
    \hrulefill
    \FOR {$n \in \{1...N\}$} 
        
        \STATE $c \leftarrow \texttt{select\_tables}(n, N)$ ~~~~~// Heuristic Rule 1 
        \STATE $p \leftarrow \texttt{Cartesian\_product}(c)$  \\
        ~~~~~~~~~~~~~~~~~~~~~~~~~~~~~~~~~~~~~~~~~~~~~~~~~~~//  Heuristic Rule 2 \& 3 
        \STATE $\texttt{solution} \leftarrow \texttt{allocate\_to\_banks}(p)$ \\
        ~~~~~~~~~~~~~~~~~~~~~~~~~~~~~~~~~~~~~~~~~~~~~~~~~~~//  Heuristic Rule 4 
        
        \IF{$\texttt{solution}$ is better than $\texttt{current\_best}$} 
            \STATE $\texttt{current\_best} \leftarrow \texttt{solution}$
        \ENDIF
    \ENDFOR

    
    \STATE \textbf{return} $\texttt{current\_best}$
\end{algorithmic}
\end{algorithm}

\section{FPGA Implementation}

In this section, we describe the implementation of MicroRec on an FPGA with an emphasis on its low inference latency.

\subsection{Reduce Latency by Deeply Pipelined Dataflow}
\label{sec:design_pipeline}

\begin{figure}[t]
  \centering
  \includegraphics[width=0.95\linewidth]{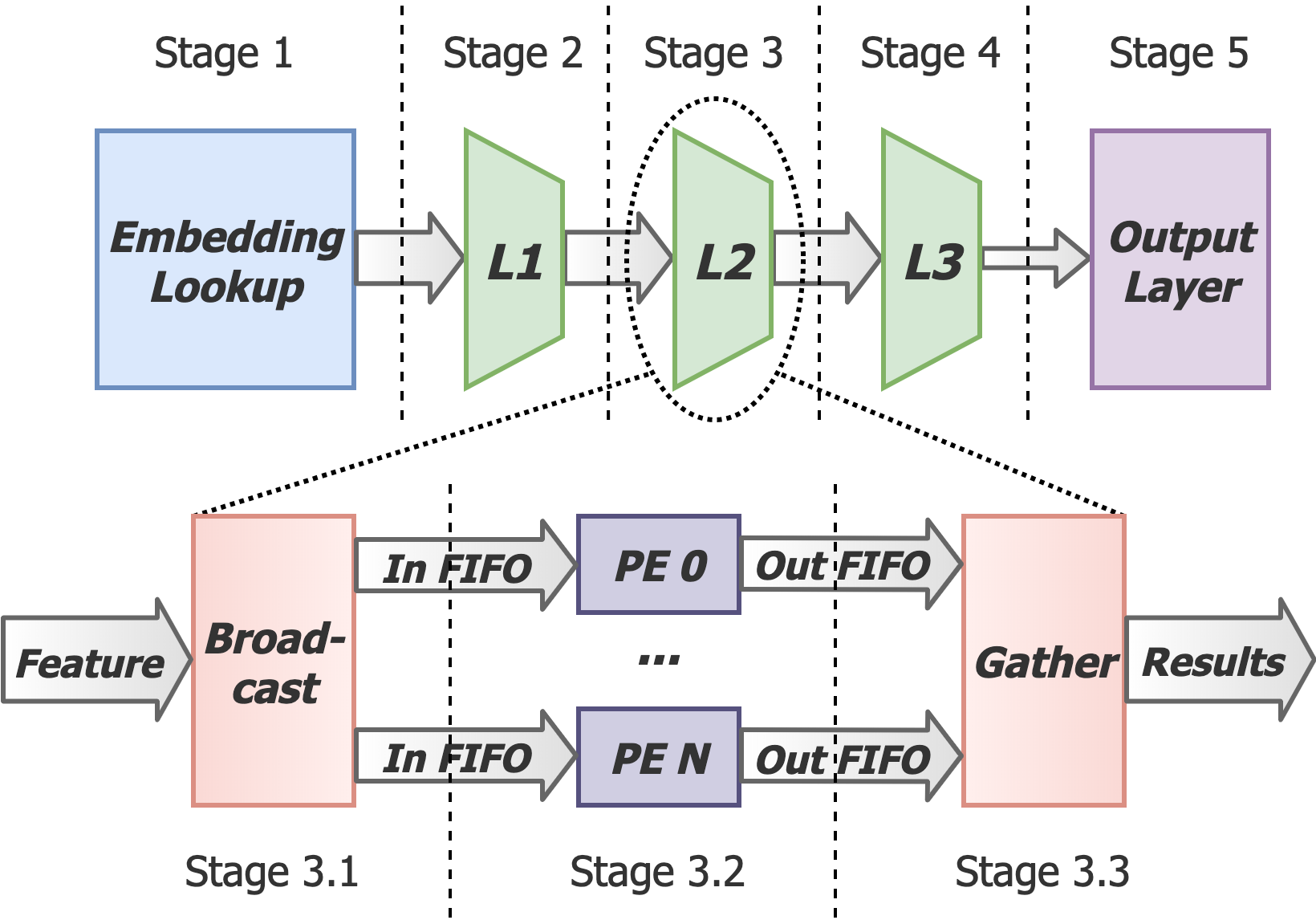}
  \vspace{-3mm}
  \caption{Highly pipelined and parallelized hardware design.}
  \vspace{-5mm}
\label{fig:FPGA_design}
\end{figure}

As shown in Figure~\ref{fig:FPGA_design}, we apply a highly pipelined accelerator architecture where multiple items are processed by the accelerator concurrently in different stages. In this design, the embedding lookup stage and three computation stages are pipelined. Each DNN computation module is further divided into three pipeline stages: feature broadcasting, computation, and result gathering. BRAMs or registers are applied to build pipes (FIFOs) as inter-module connections.

Latency concerns (SLA requirements) are eliminated by this highly pipelined design for two reasons. First, input items are processed item by item instead of batch by batch, thus the time to wait and aggregate a batch of recommendation queries is removed. Second, the end-to-end inference latency of a single item is much less than a large batch. 

\subsection{Embedding Lookup Module}
\label{sec:fpga_embedding_module}

The embedding lookup module gathers and prepares concatenated dense features for fully-connected layers. After receiving lookup indexes, the module concurrently retrieves embedding vectors from HBM, DDR, and on-chip memory banks. The concatenated embeddings are then fed to DNN computation modules through FIFOs.

\subsection{DNN Computation Module}
\label{sec:fpga_computation_module}

The lower half of Figure~\ref{fig:FPGA_design} presents the computation flow for a single FC layer, which consists of three pipeline stages: input feature broadcasting, general matrix-matrix multiplication (GEMM) computation, and result gathering. Partial GEMM is allocated to each processing unit (PE) for better routing design and potentially higher performance~\cite{johannes@matmul_hls}. Each PE conducts partial GEMM through parallelized multiplications followed by an add tree~\cite{chen2014dadiannao}.

\section{Evaluation}

We evaluate the performance of MicroRec for both end-to-end recommendation inference and embedding lookups alone. Given real-world models from Alibaba and the recent recommendation inference benchmark~\cite{facebook_benchmark}, MicroRec outperforms the optimized CPU baseline significantly under all experiment settings.

\subsection{Experiment Environment}
\label{sec:exp_hardware_platform}

We employ Xilinx Alveo U280 FPGA~\cite{xilinx_u280}, a high-end card equipped with 8GB of HBM2 (32 channels) and 32 GB of DDR4 (2 channels).
We program the FPGA by Vivado HLS~\cite{vivado_hls}, which can translate C++ programs to hardware description language (HDL).
The code is then deployed on Vitis~\cite{vitis} to generate FPGA bitstream.

The software baseline performance is tested on an AWS server with Intel Xeon E5-2686 v4 CPU @2.30GHz (16 vCPU, SIMD operations, i.e., AVX2 FMA, supported) and 128 GB DRAM (8 channels). 
We apply an open-source solution on deep recommendation systems~\cite{git_wide_deep}, where \textit{TensorFlow Serving}~\cite{olston2017tensorflow-serving, tensorflow} supports highly optimized model inference. 

\subsection{Model Specification}

We experiment the performance of MicroRec on two classes of models from different sources. The first class contains production models deployed in Alibaba, while the second class comes from the recent recommendation inference benchmark~\cite{facebook_benchmark}. 

\subsubsection{Production Models}

We experiment two deep recommendation models from Alibaba in our experiments. Both of them are memory-access intensive: they contain 47 and 98 embedding tables respectively, much more than current benchmark models~\cite{facebook_benchmark}, among which the largest model consists of only 12 tables. Table~\ref{tab:model_size} shows the parameters of our models. For example, the smaller recommendation model retrieves one vector from each of the 47 tables and gathers them into a 352-dimensional dense vector to be fed to fully-connected layers. The models we experiment do not contain bottom fully-connected layers, which are adopted in some systems to process dense input features~\cite{facebook_benchmark, ke2020recnmp}.

\begin{table}
\ra{1.2}
  \caption{Specification of the production models. }
\begin{center}
\label{tab:model_size}
\begin{small}
  \begin{tabular}{@{}  l c c c c @{}  }   
\toprule
    Model & Table Num & Feat Len & Hidden-Layer & Size \\
\midrule
    Small & 47 & 352 & (1024,512,256) & 1.3 GB \\
    Large & 98 & 876 & (1024,512,256) & 15.1 GB \\
\bottomrule
\end{tabular}
\end{small}
\end{center}
\vskip -0.1in
\end{table}

\subsubsection{Facebook Recommendation Benchmark}

We also experiment MicroRec on the recent recommendation inference benchmark by Facebook~\cite{facebook_benchmark}. The benchmark published three classes of recommendation models and their performance breakdown.  
Although we target to experiment these models for real-world-deployment, the benchmark only published a range of parameters for each type of model.
For example, the model class DLRM-RMC2 can contain from 8 to 12 tables, yet no numbers about table sizes and embedding vector lengths are provided.
Without such information, it is difficult to compare the inference performance, because some of the parameters are decisive to the inference workload. For instance, embedding vector lengths decide the number of operations to be performed in fully-connected layers.

Therefore, we compare the performance of the embedding layer: given the narrow range of table numbers \citet{facebook_benchmark} published, we can conduct multiple experiments and identify a speedup range of MicroRec.

\subsection{End-to-End Inference}
\label{sec:exp_entire}

\begin{table*}[t]  
  \caption{MicroRec performance on end-to-end recommendation inference. MicroRec achieves 2.5$\sim$5.4$\times$ speedup compared to the optimized CPU baseline (the speedup is compared to batch latency of FPGA, which consists of both the stable stages in the middle of the pipeline as well as the time overhead of starting and ending stages). Besides, the end-to-end latency to infer a single input item is as low as a couple of tens of microseconds: the latency concern of online model serving is eliminated.}
  \label{tab:exp_entire}
\begin{center}
\begin{small}
\begin{tabular}{@{} L{10em} M{3.8em} M{3.8em} M{3.8em} M{3.8em} M{3.8em} M{3.8em} M{3.8em} M{3.8em} @{}} 
\toprule
    & CPU B=1	& CPU B=64& 	CPU B=256& 	CPU B=512& 	CPU B=1024& 	CPU B=2048& 	FPGA fp16 &	FPGA fp32 \\
\midrule
    \multicolumn{9}{c}{Smaller Recommendation Model} \\
    \midrule
    Latency (ms)&	3.34 &	5.41	&8.15&	11.15	&17.17 &28.18	&\textbf{1.63E-2} & 	\textbf{2.26E-2} \\
    Throughput (GOP/s)	&0.61&	24.04&	63.81&	93.32&	121.16&	147.65&	\textbf{619.50}&	\textbf{367.72}\\
    Throughput (items/s)&	299.71 &	1.18E+4&	3.14E+4	&4.59E+4&	5.96E+4&	7.27E+4	&3.05E+5&	1.81E+5\\
    \midrule
    Speedup: FPGA fp16  &204.72$\times$  &   	24.27$\times$& 9.56$\times$& 6.59$\times$&  5.09$\times$ &\textbf{4.19$\times$}&  -&  -\\
    Speedup: FPGA fp32& 147.54$\times$&
      14.58$\times$& 5.69$\times$ &3.91$\times$& 3.02$\times$&  \textbf{2.48$\times$}& -&  - \\
    \midrule
    \multicolumn{9}{c}{Larger Recommendation Model} \\
    \midrule
    Latency (ms)&	7.48 &	10.23	&15.62	&21.06&	31.72&	56.98&	\textbf{2.26E-2} & 	\textbf{3.10E-2}\\
    Throughput (GOP/s)&	0.42&
           	19.48&	51.03&	75.66&	100.49&	111.89&	\textbf{606.41}&	\textbf{379.45}\\
    Throughput (items/s)&	133.68&	6.26E+3& 1.64E+3&	2.43E+4	&3.23E+4&	3.59E+4	&1.95E+5&	1.22E+5\\
    \midrule
    Speedup: FPGA fp16  & 331.51$\times$ &   29.56$\times$& 11.73$\times$& 7.96$\times$&  6.02$\times$ &\textbf{5.41$\times$}&  -&  -\\
    Speedup: FPGA fp32& 241.54$\times$  &  18.67$\times$& 7.36$\times$&  4.99$\times$&  3.77$\times$&  \textbf{3.39$\times$}& -&  - \\
\bottomrule
  \end{tabular}
\end{small}
\end{center}
\end{table*}

Table \ref{tab:exp_entire} compares the performance of end-to-end recommendation inference on production models between the CPU baseline and MicroRec (both Cartesian and HBM are applied). 
On the CPU side, performance increases as batch size grows, so we select a large batch size of 2048 as the baseline (larger batch sizes can break inference latency constraints). 
On the FPGA side, MicroRec infers items without batching as discussed in Section \ref{sec:design_pipeline}.
Besides, we evaluate the FPGA performance of different precision levels, i.e., 16-bit and 32-bit fixed-point numbers.

\textit{MicroRec achieves significant speedup under all experimented settings.} In terms of throughput, it is 2.5$\sim$5.4$\times$ better than the baseline under two precision levels and two model scale. Moreover, the end-to-end latency to infer a single input item is 16.3$\sim$31.0 microseconds, 3$\sim$4 orders of magnitude lower than common latency requirements (tens of milliseconds). Note that the throughput of MicroRec is not the reciprocal of latency, since multiple items are processed by the deep pipeline at the same time.

\subsection{Embedding Lookup Performance}
\label{sec:exp_embedding}

We highlight the performance boost of embedding lookups brought by Cartesian products and HBM in this section on both the production models and the benchmark models.

\subsubsection{Lookups on Production Models}
\begin{table*}[t]  
  \caption{Benefit and overhead of Cartesian products. It only costs marginal extra storage to achieve significant speedup.}
  \label{tab:cartesian_benifits}
\begin{center}
\begin{small}
  \begin{tabular}{@{} L{9em} M{5em} M{7em} M{10em} M{5em} M{8em} @{}}
\toprule
    & Table Num & Tables in DRAM & DRAM Access Rounds & Storage & Lookup Latency \\
\midrule
    \multicolumn{6}{c}{Smaller Recommendation Model} \\
    \midrule
    Without Cartesian & 47 & 39 & 2 & 100\% & 100\% \\
    With Cartesian & 42& 34& 1 & \textbf{103.2\%} & \textbf{59.2\%} \\
    \midrule
    \multicolumn{6}{c}{Larger Recommendation Model} \\
    \midrule
    Without Cartesian & 98 & 82 & 3 & 100\% & 100\% \\
    With Cartesian & 84 & 68 & 2 & \textbf{101.9\%} & \textbf{72.1\%} \\
\bottomrule
  \end{tabular}
\end{small}
\end{center}
\end{table*}

\begin{table*}[t]
\label{tab:embedding}
  \caption{MicroRec performance on the embedding layer. Given the same element data width of 32-bits, it outperformed the optimized CPU baseline by over one order of magnitude. Besides, it only took no more than one microsecond to finish lookups and concatenations even in embedding-intensive models (47 and 98 tables).}
  \label{tab:exp_embedding}
\begin{center}
\begin{small}
  \begin{tabular}{@{} L{12em} M{3.5em} M{3.5em} M{3.5em} M{3.5em} M{3.5em} M{3.5em} M{4em} M{6.5em} @{}} 
\toprule
    & CPU B=1	& CPU B=64& 		CPU B=256& 	CPU B=512& 	CPU B=1024& CPU B=2048& 	FPGA: HBM & 	FPGA: HBM + Cartesian \\
\midrule 
    \multicolumn{9}{c}{Smaller Recommendation Model} \\
    \midrule
    Latency (ms) &	2.59&	3.86&	4.71&	5.96 &8.39&	12.96&	7.74E-4&	\textbf{4.58E-4} \\

    Speedup: HBM  &3349.97$\times$ &77.91$\times$  &23.75$\times$&  15.04$\times$&  10.59$\times$&  \textbf{8.17$\times$}&  -&  - \\

    Speedup: HBM + Cartesian &  5665.07$\times$& 131.76$\times$ &40.16$\times$&  25.44$\times$& 17.91$\times$& \textbf{13.82$\times$}&  -&  - \\
    \midrule
    \multicolumn{9}{c}{Larger Recommendation Model} \\
    \midrule
    Latency (ms) &	6.25&	8.05	&10.92&	13.67&	18.11	&31.25&	1.38E-3& 	\textbf{1.03E-3} \\

    Speedup: HBM  &4531.23$\times$ &  91.29$\times$&  30.94$\times$& 19.36$\times$& 12.83$\times$& \textbf{11.07$\times$}& - &- \\

    Speedup: HBM + Cartesian &  6019.37$\times$ & 121.28$\times$&  41.10$\times$& 25.72$\times$& 17.04$\times$& \textbf{14.70$\times$} &  - &-\\
\bottomrule

  \end{tabular}
\end{small}
\end{center}
\end{table*}

\textit{MicroRec outperforms CPU baseline significantly on production models as shown in Table~\ref{tab:exp_embedding}}. Same as Section~\ref{sec:exp_entire}, a large batch size of 2048 is selected for the CPU baseline to achieve high throughput, while the accelerator always processes inputs item by item (no concept of batch sizes). 
This latency excludes streaming input features from CPU side memory as mentioned in footnote \ref{foot_CPU_input}. The result shows that MicroRec outperforms the baseline by 13.8$\sim$14.7$\times$  on the embedding layer (in addition to DRAM accesses, the many embedding-related operator calls in TensorFlow also leads to large consumption in the CPU baseline). 
Some detailed result interpretation includes:


\textit{Though HBM can achieve satisfying performance on its own, Cartesian products further speed up the process.} For the smaller model, as shown in Table~\ref{tab:cartesian_benifits}, except those tiny tables stored on-chip, there are still 39 tables left to be allocated to DRAM. Considering there are 34 DRAM channels in total (32 for HBM, 2 for DDR), it takes two DRAM access rounds to lookup 39 tables. Cartesian products can reduce the table number to 34, so that only one round of DRAM access is required. The experiment shows that, with Cartesian products, the latency of embedding lookup is only 59.17\% of the HBM-only solution (458 ns vs 774 ns). Similarly, for the larger model, Cartesian products reduce the memory access rounds from 3 to 2, consumed only 72.12\% of the time (1.63 us vs 2.26 us).

\textit{The storage overhead of Cartesian products is fairly low.} As shown in table \ref{tab:cartesian_benifits}, the products only lead to 3.2\% and 1.9\% storage overhead on the two models respectively. This is because only small tables are selected for Cartesian products as introduced in section \ref{sec:microrec_algorithm}, so that the products are still considerably small compared to a single large table.

\textit{By Cartesian products and HBM, the memory bottleneck caused by embedding lookup is eliminated.} Since the embedding lookups only cost less than 1 microsecond in MicroRec (as in Table~\ref{tab:exp_embedding}), the bottleneck shifts back to computation, in which the most expensive stage takes several microseconds. 

\begin{figure}
\centering
\includegraphics[width=0.95\linewidth]{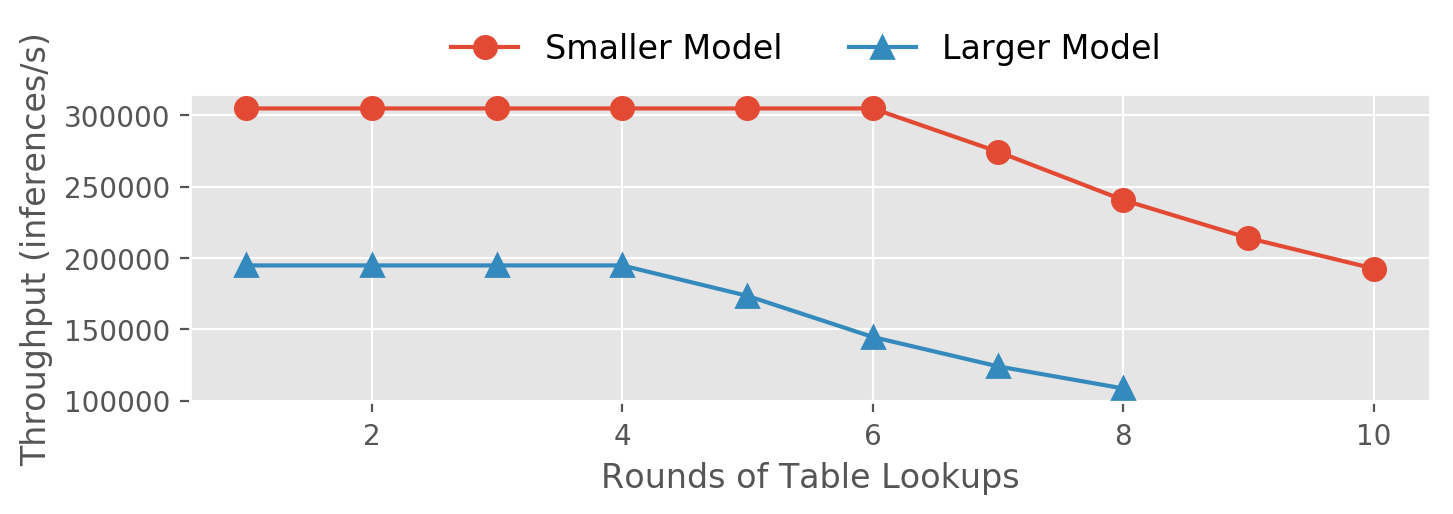}
\vspace{-1em} 
\caption{End-to-end inference throughput of MicroRec. It allows multi-rounds lookup without sacrificing performance.}
\label{fig:multi-round-lookups}
\end{figure}

\textit{The accelerator performance is robust even as multiple rounds of lookups are required.} Although the production models only involves one lookup operations per table, alternative DNN architectures may require multiple rounds of lookups~\cite{facebook_benchmark}. Figure~\ref{fig:multi-round-lookups} proves the performance robustness of MicroRec in such scenarios by assuming more rounds of embedding retrievals on the two production models --- the smaller and larger models can tolerate 6 and 4 rounds of lookups without downgrading the end-to-end inference throughput at all using 16-bit fixed-points, because the DNN computation and embedding lookup stages are overlapped. Once more rounds of lookups are assumed, the performance starts to depend on the total memory access latency which is proportional to the rounds of DRAM accesses.

\subsubsection{Performance on Benchmark Models}

\begin{table}
  \caption{MicroRec achieves 18.7$\sim$72.4$\times$ embedding lookup speedup compared to the Facebook's recommendation baseline.}
  \label{tab:facebook_benchmark}
\begin{center}
\begin{small}
  \begin{tabular}{@{} L{6em} c c c c c @{}} 
\toprule
\multirow{2}{*}{Performance} & \multicolumn{5}{c}{Embedding Vector Length} \\
\cmidrule{2-6}
     & 4	& 8 & 16 & 	32 & 64  \\
\midrule 
\multicolumn{6}{c}{8 Tables (Speedup Upper Bound)}\\
\midrule
\vspace{0.2em} Lookup (ns) &	334.5 & 353.7 & 411.6 &  486.3 & 648.4 \\
\vspace{0.2em} Speedup & 72.4$\times$ & 68.4$\times$ & 58.8$\times$ & 49.7$\times$ & 37.3$\times$ \\
\midrule
\multicolumn{6}{c}{12 Tables (Speedup Lower Bound)}\\
\midrule
\vspace{0.2em} Lookup (ns) & 648.5 & 707.4 & 817.4 & 972.7 & 1296.9 \\
\vspace{0.2em} Speedup & 37.3$\times$ & 34.2$\times$ & 29.6$\times$ & 24.8$\times$ & 18.7$\times$ \\
\bottomrule

  \end{tabular}
\end{small}
\end{center}
\end{table}

We compare the embedding lookup performance of MicroRec to the recent recommendation inference benchmark~\cite{facebook_benchmark}. Although the paper does not expose all model parameters, we can still identify the embedding lookup performance range on MicroRec by experimenting a range of table settings.
To be more specific, we experiment the embedding-dominated model class DLRM-RMC2, which contains 8$\sim$12 small tables and each table is looked up 4 times (thus $32\sim48$ lookups in total). 
Several assumptions are made for the missed information. First, by ``small tables'', we assume each table is within the capacity of an HBM bank (256MB). Second, we assume common embedding vector lengths from 4 to 64. Third, no Cartesian products are applied in our experiments, since the table sizes are assumed by us. 


Table \ref{tab:facebook_benchmark} shows the embedding lookup performance on MicroRec: it achieves 18.7$\sim$72.4$\times$ speedup compared to the published baseline performance (2 sockets of Broadwell CPU @2.4GHz; 14 cores per socket; AVX-2 supported; 256 GB 2400MHz DDR4 DRAM; batch size=256). This performance range is identified by experimenting table numbers from 8 to 12 and vector lengths from 4 to 64. The highest speedup occurred when there are only 8 embedding tables (32 lookups) with a short vector size of 4, for which only one round on HBM lookup is required.
The lowest speedup happens when there are 12 tables with a long vector size of 64, where 2 rounds of HBM accesses are necessary.  

\section{Related Work}
\label{sec:related_work}

\textit{Deep learning for personalized recommendations.} While regular DNNs take dense features as input, \citet{he2017neural} proposed to encode user information by embedding tables in recommendation models. The dense features gathered from embedding vectors are then processed by a series of matrix-factorization and FC layers. \citet{deep_youtube} applied similar neural network structures for Youtube video recommendation. \citet{wide_and_deep_app_store} discussed the pros-and-cons of linear models and deep models, and proposed to serve Google Play recommendations with joint wide-and-deep models. \citet{wide_and_deep_MT_youtube} improved the wide-and-deep model by taking multiple objectives into account beyond click-through rates (CTR), e.g., comments, likes, and ratings. Facebook introduced additional fully-connected layers to process dense input features: the dense features are fed into bottom FC layers, and the output features are concatenated with embedding vectors~\cite{facebook_benchmark}. Alibaba removed dense input features in their deep models and applied attention mechanism on the top of embedding tables~\cite{din_alibaba_attention_fc}. \citet{dien_alibaba_attention_rnn} further extended this work by introducing sequential neural network.


\textit{Hardware solutions for recommendations.}
According to Facebook, recommendation workloads can consume up to 79\% of total AI inference cycles in data centers~\cite{facebook_benchmark}.
However, little research has been focused on serving personalized recommendations efficiently. In order to provide enough background knowledge to the research community and tackle this important problem, \citet{facebook_benchmark} analyzed the recommendation workload comprehensively, open-sourced several models used by Facebook, and set up a performance benchmark.
\citet{tensordimm_kaist} is the first hardware solution for high performance recommendation inference. They reduced the memory bottleneck by introducing DIMM-level parallelism in DRAM and supporting tensor operations, e.g., gather and reduction, within the DRAM. \citet{ke2020recnmp} extended the idea of near-memory-processing and added memory-side-caching for frequently-accessed entries. \citet{gupta2020deeprecsys} took into account the characteristics of query sizes and arrival patterns, and developed an efficient scheduling algorithm to maximize throughput under latency constraints by using both CPUs and GPUs. \citet{hwang2020centaur} implemented an FPGA accelerator (without HBM) for deep recommendation inference, and the speedup was significant for models with few embedding tables. Compared to previous work, MicroRec is the first system that introduces data structure solution, i.e., Cartesian products, to reduce the number of DRAM accesses. It is also the first work resorting to HBM so as to parallelize embedding lookups.

\textit{Efficient Model Serving.} Aside from recommendations, a range of research has been focused on neural network serving. Due to the heavy workload of DNN inference, many works resort to specialized hardware~\cite{jouppi2017datacenter, mei2019sub, chung2018serving, hsieh2018focus, zhang2017improving, chen2016eyeriss, mao2019mobieye, sharify2019laconic, shao2019simba, feng2019asv, owaida2017scalable, gao2017tetris, chen2014dadiannao, hua2019boosting, farcas2020hardware}. 
Besides, designing hardware-efficient neural networks is essential for inference performance~\cite{han2017ese, stamoulis2019single, teja2018hydranets, zhang2019skynet, howard2017mobilenets, elthakeb2018releq, ghasemzadeh2018rebnet, maschi2020making}. 
Furthermore, one can optimize serving performance on general-purpose hardware (CPU and GPU) by system-level optimization~\cite{olston2017tensorflow-serving, narayanan2018accelerating, chen2018tvm, choi2020prema, wu2019machine, crankshaw2018inferline}.

\textit{Efficient Model Training.} Due to the increasing numbers and sizes of neural networks, high-performance model training becomes essential~\cite{mattson2019mlperf}. Training usually resorts to accelerators such as GPUs~\cite{shoeybi2019megatron, cho2019blueconnect, dong2020eflops, cui2016geeps} and FPGAs~\cite{zhang2017zipml, cho2019fa3c,kara2017fpga,he2020km,he2018flexible,zhao2016f, gurel2020compressive}. Besides, many works accelerate training by better system and algorithm designs~\cite{jayarajan2019priority, das2018mixed, peng2019generic, narayanan2019pipedream, jia2018beyond, wang2019accelerating, moritz2018ray, kurth2017deep, rajbhandari2017optimizing, li2020halo, li2014scaling, chen2018effect, abuzaid2016yggdrasil}.

\section{Conclusion}

We design and implement MicroRec, a high-performance deep recommendation inference engine. On the data structure side, MicroRec applies Cartesian products to reduce sparse memory accesses. On the hardware side, HBM is adopted to scale up embedding lookup concurrency, and the deeply pipelined architecture design on FPGA enables low inference latency. 
By the three strategies we propose, the memory bottleneck caused by embedding lookups is almost eliminated, and the latency requirements of recommendation inference are easily met.

\section*{Acknowledgements}
Part of the work of Wenqi Jiang and Zhenhao He has been funded by the Alibaba Group. We would like to thank Xilinx for their generous donation of the XACC FPGA cluster at ETH Zurich on which the experiments were conducted.

\bibliographystyle{mlsys2021}
\bibliography{ref}


\newpage
\appendix
\onecolumn{
\section*{APPENDIX: MEMORY CONTROLLER AND AXI INTERFACE}
\vspace{3mm}

To set up the communication between FPGA and DRAM (including HBM and DDR), we choose a narrow AXI interface data width of 32-bit. Although the full data width (512-bit) can reduce the number of clock cycles required for vector reading, it has two disadvantages. First, it consumes too much hardware resources. To support efficient communication to DRAM without much stalls, we apply BRAMs as long FIFOs. Since there are 34 DRAM channels in total (32 for HBM and 2 for DDR), these FIFOs will consume over half of total BRAMs slices on Alveo U280 FPGA given 512-bit data width. Such BRAM consumption is too expensive to afford because DNN computation modules also require substantial BRAM resources. Second, higher resource utilization can lead to downgraded clock frequency, resulting in lower inference performance. According to the experiments in section~\ref{sec:exp_entire} and~\ref{sec:exp_embedding}, the embedding lookup process in our design is fast enough to be covered by DNN computation (remember we applied a pipelined design). As a result, lower clock frequency will lead to decreased computation performance thus higher inference latency.


\section*{APPENDIX: FPGA RESOURCE UTILIZATION}


\vspace{3mm}

Table \ref{tab:resource_util} lists the resource utilization and clock frequency of our deep recommendation inference accelerator. We implement the design on Xilinx Alveo U280, a high-end FPGA card consisting of three die areas. 
The resource consumptions are composed of all GEMM PEs, their interconnection, and the embedding lookup module.
According to the estimation of Vivado HLS (the consumption can be further optimized by the Vivado backend), each PE for 32-bits fixed-point GEMM consumes 7 BRAM slices and 18 DSPs while the 16-bit one consumes 4 BRAM slices and 14 DSPs. The number of PEs for three layers are 128, 128, and 32 for both models and precision-levels.
Because of the high resource utilization rate (more than 80\% for some resources), cross-die routing is necessary, and the long-distance communication must be tolerated by low clock frequency (120$\sim$140MHz).

\begin{table}[h!]  
  \caption{FPGA frequency \& resource utilization (Xilinx Alveo U280)}
  \vspace{1em}
  \begin{tabular}{ M{8em} M{8em} M{8em} M{8em} M{8em}  }   
    \hline
     & \multicolumn{2}{c}{Small Model} & \multicolumn{2}{c}{Large Model} \\
    \hline
    Precision & fixed-point 16 & fixed-point 32 & fixed-point 16 & fixed-point 32 \\
    Freq (MHz) & 120 & 140 & 120 & 135 \\
    \hline
    \multicolumn{5}{c}{Utilization (Slices)} \\
    \hline
    BRAM 18Kbit & 1,566 & 1,657 & 1,566 & 1721 \\
    DSP48E & 4,625 & 5,193 & 4,625 & 5,193 \\
    Flip-Flop & 683,641 & 764,067 & 691,042 & 777,527  \\
    LUT & 485,323 & 568,864 & 514,517 & 584,220 \\
    URAM 288Kbit & 642 & 770 & 642 & 770 \\
    \hline
    \multicolumn{5}{c}{Utilization (\%)} \\
    \hline
    BRAM 18Kbit & 78 & 82 & 78 & 85 \\
    DSP48E & 51 & 57 & 51 & 57 \\
    Flip-Flop & 26 & 29 & 27 & 30 \\
    LUT & 37 & 44 & 40 & 45 \\
    URAM 288Kbit & 66 & 66 & 80 & 80 \\
    \hline

  \end{tabular}
  \label{tab:resource_util}
\end{table}

\section*{APPENDIX: COST ESTIMATION}
\vspace{3mm}

We compare the price between CPU-based and FPGA-based inference engine on AWS. The CPU server we rent costs \$1.82 per hour while renting an FPGA server only costs \$1.65 (AWS provides U250, a similar model to what we use). Considering the 4$\sim$5x speedup using 32-bit fixed-points, deploying FPGAs will be beneficial in the long-term.

}

\end{document}